# An Illumination- and Temperature-Dependent Analytical Model for Copper Indium Gallium Diselenide (CIGS) Solar Cells

Xingshu Sun,[1] Timothy Silverman,[2] Rebekah Garris,[2] Chris Deline,[2] and Muhammad Ashraful Alam[1]

[1]School of Electrical and Computer Engineering, Purdue University, West Lafayette, Indiana, USA

[2] National Renewable Energy Laboratory, Golden, Colorado, 80401, USA

*Abstract* — **In this paper, we present a physics-based analytical model for CIGS solar cells that describes the illumination- and temperature-dependent current-voltage (I-V) characteristics and accounts for the statistical shunt variation of each cell. The model is derived by solving the drift-diffusion transport equation so that its parameters are physical, and, therefore, can be obtained from independent characterization experiments. The model is validated against CIGS I-V characteristics as a function of temperature and illumination intensity. This physics-based model can be integrated into a large-scale simulation framework to optimize the performance of solar modules as well as predict the long-term output yields of photovoltaic farms under different environmental conditions.**

## I. Introduction

Among the commercially available photovoltaic (PV) technologies (e.g., c-Si, CdTe, CIGS), CIGS solar cells are attractive because of their moderately high efficiency with low-cost production (e.g., less material usage and low-temperature processes). Device performance has improved steadily, and power conversion efficiency exceeding 22.3% has been reported at the cell level [1]. The performance gain has not been fully reflected at the module level (current highest module efficiency ~17% [2]). Therefore, there are opportunities for manufacturing and design improvements to reduce the cell-to-module efficiency gap [3]–[5].

Toward this goal, a SPICE-based large-scale simulation framework can help analyze and optimize the performance and reliability of large-area solar modules [6]–[8]. The framework consists of a network of thousands of cell-level equivalent circuits connected in parallel and series to represent a solar module [9]. In this context, an analytical (compact) model that can describe the current-voltage (I-V) characteristics and the statistical variation of small-area cells, is indispensable. Such a module-level simulation framework can be leveraged for system-level studies to predict long-term energy yields for solar farms. Because the outside environment (e.g., temperature, solar irradiance, shadow) is specific to geographical locations and changes spatially and temporally, the analytical model must therefore reflect illumination- and temperature- dependencies of cell performance.

Solar cells can be categorized into three groups depending on their device configurations: **p-n homojunction** (e.g., c-Si, GaAs), **p-i-n junction** (e.g., a-Si, perovskites), or **p-n heterojunction** (e.g., CIGS, CdTe). Because ideal p-n homojunction solar cells obey the superposition principle (*i.e.*, its light current can be represented as the uncorrelated sum of its diode dark current and a constant photocurrent), the conventional 5-parameter circuit model (consisting of a diode, constant current source, linear shunt resistor, and external series resistor) describes its performance well [10]. In contrast, p-i-n junction and p-n heterojunction solar cells do not follow the superposition principle; therefore, the 5-parameter model cannot accurately describe their I-V characteristics [11]. Recently, we have reported on a physics-based model for p-i-n cells [12], [13]. In this paper, we complete the set by reporting on a physics-based model for p-n heterojunction cells, and specifically, CIGS solar cells.

Given the commercial importance of CIGS cells, researchers have long recognized the failure of superposition [14] and the importance of voltage-dependent carrier collection [15]–[17] in these p-n heterojunction cells. Unfortunately, existing analytical models do not explicitly capture illumination or temperature dependencies of CIGS cells, because the models account for only one of the several factors responsible for voltage-dependent collection (e.g., [18] interprets the voltage-dependent depletion, whereas [19] focuses on interface recombination). Moreover, the model in [18] was originally derived for a reverse-biased p-n junction [20], which may not accurately predict forward-biased characteristics of a solar cell. Also, the nonlinear voltage-dependence and the statistical distribution of shunt currents contribute to efficiency losses in monolithic CIGS modules, and their effects have not been modeled quantitatively.

In this paper, we present a physics-based analytical model for CIGS solar cells that explicitly accounts for non-uniform photo-generation, conduction and valence band offsets, generation-dependent carrier recombination, bias-dependent depletion, and log-normal distributed shunt leakage current. This physics-based model anticipates the irradiance- and temperature-dependent I-V characteristics observed in experiments. The model equations have been applied in [6] to specifically address the problem of shadow degradation in CIGS modules. This paper provides a full derivation of the model equations, explains assumptions explicitly, and provides additional details needed to tailor this model to specific applications.

## II. FORWARD I-V OF A HETEROSTRUCTURE CELL

In general, the light I-V curve of a solar cell can be written as:

$$J_{Light}(IL,V) = J_{Photo}(IL,V) - J_{Diode}(IL,V), \quad (1)$$

where $J_{Diode}$ is the diode current (also referred to as the injection current $J_{Inj}$ in [21]) due to the carriers injected from the contacts, and $J_{Photo}$ is the photocurrent due to the photo-generated carriers. These two current components can depend on illumination intensity ($IL$) and applied voltage ($V$); in other words, Eq. (1) does not presume superposition. To obtain the analytical solution of the full I-V characteristics for heterojunction structures, we solve for photocurrent and diode current separately. The final equations are summarized in Table A1 and are derived below. The analytical results are validated against experimental data as well as detailed self-consistent optoelectronic simulations using the commercial device simulator Sentaurus [22]. The physical parameters are taken from [23]; see Table A2. Appendix B summarizes the simulation results and discusses the validity of various assumptions.

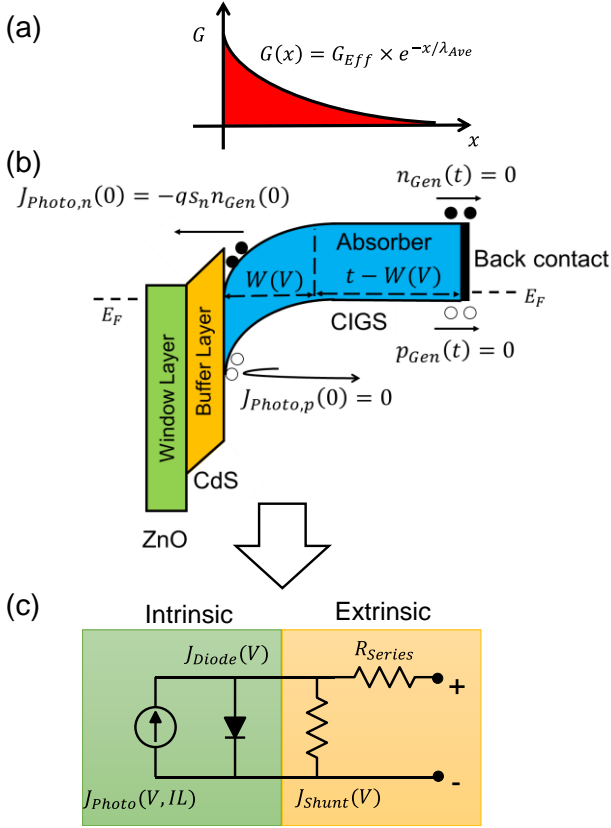

Fig. 1 (a) The exponential generation profile in the CIGS absorber layer. (b) The energy band diagram of a typical CIGS cell. The boundary conditions labeled here are only for the generated carriers. (c) The equivalent circuit diagram for CIGS solar cells. The parameters and analytical equations for each element are summarized in Tables A1 to A2.

### A. Photocurrent ($J_{Photo}$)

A typical CIGS cell consists of a ZnO window layer (~200 nm), an ultrathin (~50 nm) but large-bandgap CdS buffer layer, stacked on top of a thick (1~3 μm) CIGS absorber layer [23]; see Fig. 1(b). $J_{Photo}$ can be obtained analytically by solving the position-dependent continuity equations *only for the photo-generated carriers* [24], [25], namely,

$$D\frac{\partial^2 n_{Gen}(p_{gen})}{\partial x^2} + \mu E(x)\frac{\partial n_{Gen}(p_{Gen})}{\partial x} + G(x) - R_{Gen}(x) = 0, \quad (2)$$

$$J_{Photo,n(p)} = q\mu E(x)n_{Gen}(p_{Gen}) \pm qD\frac{\partial n_{Gen}(p_{Gen})}{\partial x}. \quad (3)$$

Here, $n_{Gen}(p_{Gen})$ is the generated electron (hole) concentration; $D$ and $\mu$ are the diffusion coefficient and mobility, respectively; $G(x)$ represents the *position-resolved generation*, as in Fig. 1(a); and $R_{Gen}(x)$ is the bulk (radiative, Auger, Shockley-Read-Hall) recombination of the photo-generated carriers before they are collected by the contacts. Note that $R_{Gen}(x)$ is the difference of bulk recombination rates under light and dark conditions ($R_{Gen}(x) = R_{Light}(x) - R_{Dark}(x)$). Finally, $E(x)$ is the position-dependent electric field within the absorber layer.

Equations (2) and (3) are coupled nonlinear equations, amenable only to numerical solutions, as with ADEPT [26] or Sentaurus [22]. *However, with two simplifications related to $G(x)/R_{gen}(x)$ and the field-dependent carrier collection, we can solve the equation analytically, as follows.*

Approximation 1: Recombination-Corrected Generation

Neglecting parasitic generation in the window and buffer layers and reflectance from the back metal contact, Beer's law allows us to approximate the generated profile in the absorber layer as $G(x) = G_{Eff}e^{-x/\lambda_{Ave}}$, where $G_{Eff}$ and $\lambda_{Ave}$ are the material specific constants, averaged over the solar spectrum. Therefore, the total photocurrent in the absence of bulk recombination is $J_{Tot-Photo} = q\int_0^\infty G_{Eff}e^{-x/\lambda_{Ave}}dx = qG_{Eff}\lambda_{Ave}$. The recombination term, $R_{Gen}(x)$, is determined by $n_{Gen}(x)$ and $p_{Gen}(x)$ as well as the carrier lifetime. Using self-consistent optoelectronic numerical simulation, we find that the generation-induced bulk recombination current ($J_{Gen-Rec} = \int_0^t R_{Gen}(x)dx$) is *voltage-independent and remains a small fraction of the total photocurrent up to open-circuit voltage ($V_{OC}$)*, see Fig. A4(a). This is because, for $V < V_{OC}$, $R_{Gen}$ occurs primarily in the quasi-neutral bulk region, where the bulk recombination is voltage-independent; see Fig. A3. In addition, $J_{Gen-Rec}$ scales linearly with $J_{Tot-Photo}$ under different illumination intensities ($J_{Gen-Rec} = a \times J_{Tot-Photo}, a \approx 5\%$) as shown in Fig. A4(b). Therefore, we can account for $R_{Gen}(x)$ in Eq. (2) by normalizing the generation profile to $G'(x) = G_{Eff}'e^{-x/\lambda_{Ave}}$ so that the short-circuit current, $J_{SC} = J_{Tot-Photo} - J_{Gen-Rec} = qG_{Eff}'\lambda_{Ave}$. Consequently, Eq. (2) can be rewritten as

$$D \frac{\partial^2 n_{Gen}(p_{Gen})}{\partial x^2} + \mu E(x) \frac{\partial n_{Gen}(p_{Gen})}{\partial x} + G'(x) = 0. \quad (4)$$

Approximation 2: Linearity of the Depletion Field

We can further simplify Eq. (4) by carefully analyzing the electric field, $E(x)$, from numerical simulation. Specifically, numerical simulation shows that within the p-type absorber, $E(x)$ is linear inside the depletion region ($x \leq W$), i.e., $E(x) = [1 - x/W(V)] E_{Max}(V)$, but vanishes beyond it ($x > W(V)$). Here, $W$ is the depletion width and $|E_{Max}(V)| = 2\beta(V_{Bi} - V)/W(V)$; $V_{Bi}$ and $V$ are respectively the total built-in potential and the bias voltage across both n and p sides of the junction. Since we only consider $E(x)$ within the p-type absorber, a parameter $\beta$ is introduced to account for the voltage partition between absorber and window layers, exactly analogous to voltage partition between channel and the oxide in a metal oxide semiconductor (MOS) capacitor [27]. Specifically, with applied voltage $V$ for the entire device, a $\beta \times V$ drops across the absorber layer, while $(1 - \beta) \times V$ drops across the window and buffer layers. Our detailed numerical simulation shows that photo-generation up to one-sun illumination does not significantly perturb the electric field; therefore, $E(x)$ in Eq. (4) is assumed to be independent of illumination intensity.

Flux Boundary Conditions

Finally, one needs to set the boundary conditions to solve Eqs. (3) and (4). *For electrons*, the photocurrent at the interface between the buffer and absorber layers (see Fig. 1(b)) is given by $J_{Photo,n} = -q \times s_n \times n_{Gen}$, where $s_n = v_R e^{-\Delta E_C/kT}$ defines the interface thermionic-emission velocity and $\Delta E_C$ is the conduction band offset and $v_R$ is the Richardson velocity [25]; large $\Delta E_C$ reduces the thermionic velocity; the reduced carrier-collection efficiency distorts the I-V characteristics [28]. *For holes*, the valence band offset $\Delta E_V$ at the buffer/absorber interface is presumed to be large enough so that $J_p = 0$; see Fig. 1(b). The back contact at $x = t$ is treated as an ideal ohmic contact (negligible Schottky barrier) for both electrons and holes, so that $n_{Gen} = p_{Gen} = 0$, as shown in Fig. 1(b). The assumption of a Schottky-barrier-free back contact is supported by experimental observations [29]–[31]. If needed, a back-to-back diode circuit can be added to this model to account for the Schottky barrier [32].

Analytical Solution of Eqs. (3) and (4).

Integrating $G'(x)$, $E(x)$, two aforementioned assumptions, and the flux boundary conditions into Eqs. (3) and (4), we obtain the solution of the photocurrent as

$$J_{Photo} = J_{SC} \times f_{Opt} \times f_{Coll}. \quad (5)$$

$$f_{Opt} \equiv 1 - \frac{\lambda_{Ave}}{W(V)} (\exp(-\frac{t-W(V)}{\lambda_{Ave}}) - \exp(-\frac{t}{\lambda_{Ave}})), \quad (6)$$

$$f_{Coll} \equiv (1 + \frac{v_{Diff}}{s_n \times \exp(\frac{q\beta(V_{Bi}-V)}{kT})})^{-1}. \quad (7)$$

The parameters in Eqs. (5)–(7) are physical and can be calibrated using independent measurements. For example, $t$ is the thickness of the CIGS layer, and $W(V)$ and $V_{Bi}$ can be estimated from the Mott-Schottky analysis of capacitance-voltage (C-V) measurements.

Despite their apparent complexity, Eqs. (5)–(7) can be explained in simple terms. For example, Eq. (6) determines the total optical absorption in the solar cell. So if the cell thickness $t$ is on the order of $\lambda_{Ave}$ ($t \approx \lambda_{Ave}$), then $f_{Opt} < 1$, indicating incomplete absorption of photons. The formulation of Eq. (6) also ensures that the optical efficiency, $f_{Opt}$, is bounded between 0 and 1 for any combination of $\lambda_{Ave}$, $W(V)$, and $t$ ($> W(V)$). Similarly, Eq. 7 governs the efficiency of carrier collection $f_{Coll}$. In Eq. (7), $v_{Diff} \equiv \frac{D}{t - W_{Delp}(V)}$ is the diffusion velocity in the quasi-neutral region and $s_n \equiv v_R e^{-\Delta E_C/kT}$ is the interface thermionic-emission velocity for electrons at the heterojunction. Hence, Eq. (7) balances two competing transport processes of the photogenerated electrons: 1) field-assisted drift toward the heterojunction interface followed by thermionic emission to the front contact, and 2) back-diffusion through the quasi-neutral region followed by recombination at the "wrong" (back) contact. Increasing forward-bias $V$ reduces the depletion field, resulting in the increasing fraction of electrons diffusing to the back contact and recombining there instead of contributing to photocurrent; correspondingly, the collection efficiency, $f_{Coll}$, decreases. Recall that the carrier loss due to bulk recombination is explicitly accounted for in the normalized generation profile in Eq. (4), which assumes $t \gg W(V)$.

For practical CIGS cells [23], $t \gg \lambda_{Ave}$, $t \gg W(V)$, and $\eta_{Opt} \to 1$. With these approximations, the photocurrent simplifies to

$$J_{Photo} \simeq J_{SC} \frac{1}{1 + \alpha_C \times \exp(\frac{q\beta(V_{Bi}-V)}{kT})}. \quad (8)$$

Here, $\alpha_C \approx \frac{D}{t \times s_n} \equiv \frac{v_{Diff}}{s_n}$, which is the ratio between diffusion velocity and thermionic-emission velocity. Equation (8) implies that, for high-quality CIGS cells, the heterojunction (accounted in $\alpha_C$) is the main cause for voltage-dependent carrier collection, which is in agreement with [16], [33]. Equation (8) is a simplified version of Eqs. (5)–(7) preferred for use in large-scale module simulation for numerical speed and robustness.

It is interesting to compare Eq. (8) to previously published equations (Eq. (2) in [18] and Eq. (3) in [34]), i.e., $J_{Photo} \simeq J_{SC}(1 - \frac{\exp(-W_{Delp}(V)/\lambda_{Ave})}{L_{Diff}/\lambda_{Ave}+1})$, where $L_{Diff}$ is the carrier diffusion length. The equation implies that if $L_{Diff} \to \infty$ (no bulk recombination), then $J_{Photo}$ equals $J_{SC}$, independent of bias voltage. The detailed simulation in [16], however, shows that even in the absence of bulk recombination, $J_{Photo}$ in heterojunction devices should be zero at $V = V_{Bi}$ due to carrier partition, consistent with Eq. (8) ($J_{Photo} \simeq 0$ at $V = V_{Bi}$ given $\alpha_C \gg 1$). Hence, Eq. (8) is an improvement on the equation in [18], [34], because the physics of carrier partition between drift and diffusion in the forward bias was captured.

Let us consider the temperature and intensity dependencies of $J_{Photo}(IL,T)$. Among the four parameters of the photocurrent in Eq. (8), **first**, $J_{sc}$ is proportional to $IL$, but is essentially independent of $T$ because the bandgap of CIGS is not temperature sensitive [35]. Our measurements support this assertion (see Sec. III): $J_{sc}$ of our CIGS solar samples is measured to be temperature independent within the temperature range of interest (260 K to 360 K), which agrees with [36], [37]. **Second**, the $T$ dependency of $V_{Bi}$ of a heterojunction can be analytically described in terms of the bandgap, $E_G$, and the conduction band offset, $\Delta E_C$; see A(9) [38]. **Third**, the $T$ dependency of $\alpha_c$ in Eq. (8) can be approximated as $\alpha_c \sim e^{\Delta E_C/kT}$, because $\alpha_c$ is proportional to $1/s_n$ and $s_n \sim e^{-\Delta E_C/kT}$. **Fourth**, the voltage partition factor $\beta$ is assumed to be temperature independent, because doping density does not change significantly in the temperature range of interest. Note that the temperature and irradiance dependences of photocurrent in Eq. (8) are handled implicitly through the dependence of its underlying variables.

### B. *Diode Current ($J_{Diode}$)*

One also has to obtain the analytical description of $J_{Diode}$ in Eq. (1) to complete the model. The complexity of the diode current depends on the solar cell type. For example, the 5-parameter model shows that diode current in a p-n junction cell can be described by two exponential terms with ideality factors 1 and 2, respectively [10]. The dark currents in p-i-n cells (e.g., perovskite, a-Si) are more complex, but can nonetheless be derived analytically [8], [12], [13]. Although a similar approach can be used to derive the diode current for heterojunction cells (e.g., HIT [33]), two considerations simplify the problem significantly. *First*, numerical simulation shows that the diode current is independent of illumination (up to 1 sun; see Fig. A1), which has been supported by our experimental results (see Fig. A2). *Second*, numerical simulation and experimental data for a variety of cells also show that the voltage dependence of the diode current can be expressed as

$$J_{Diode} = J_0(T)(\exp(\frac{qV}{N(T)kT}) - 1). \quad (10)$$

Here, $J_0(T)$ is the temperature-dependent reverse saturation current, which is directly related to the bandgap and carrier diffusion length of the absorber; $N$ is the ideality factor ranging from 1–2, depending on the distribution of defects.

Regarding the two parameters ($J_0$ and $N$) of the diode current in Eq. (10), there have been extensive studies regarding their temperature dependencies [39], [40]. It has been argued that $J_0$ is linear with $e^{-E_G/N(T)kT}$ and $1/N(T) = 1/2 \times (1 + T/T^*)$ (assuming no tunneling), where $E_G$ is the absorber bandgap, and $kT^*$ (ranging from 30 meV to 150 meV) is the characteristic slope of the exponentially distributed defects in the absorber. Large $kT^*$ indicates that most of the recombination in the depletion region is through the mid-bandgap defects, giving an ideality factor $N = 2$. Small $kT^*$ corresponds to significant density of states of the shallow-level defects close to the valence band; carrier recombination due to such shallow defects gives $N = 1$.

### C. *Series and Shunt Resistances*

Series Resistance. Finally, as shown in Fig. 1(c), one must specify shunt and series resistances to complete the model, because they contribute parasitic power loss and increase the cell-to-module efficiency gap. The series resistance depends on the resistivity and thickness of the transparent conductive oxide (TCO) as well as the geometry of the cell [41] and the bulk resistivity of the absorber. It has been shown in [42] that the resistive loss due to 3-D current flow through the TCO layer and the bulk absorber in a solar cell can be modelled by a single resistor, $R_{series}$. Hence, $R_{series}$ in Fig. 1(c) can be easily specified from measurements [43].

Nonlinearity of Shunt Conduction. Next, let us consider the shunt current. In Si cells, $R_{shunt}$ is modeled as a symmetric linear resistor. For thin-film solar cells, shunt current is also symmetric, but conduction is typically non-ohmic due to space-charge-limited (SCL) transport across the absorber layer [44], [45]. A careful examination of the experimental data [45] of CIGS cells shows that transport transitions from linear to nonlinear shunt current with increasing voltage, namely,

$$J_{Shunt} = G_{Shunt} \times V + I_{OShunt} \times V^\gamma, \quad (11)$$

where $G_{Shunt}$ and $I_{OShunt}$ are the prefactor of the linear and nonlinear shunt current, respectively, and $\gamma$ is the power index of SCL transport determined by the defect distribution in the absorber. It has been shown experimentally that $J_{shunt}$ depends weakly on temperature and illumination [45], so we do not consider it explicitly. Note that Eq. (11) has been applied widely to analyze the performance limit of CIGS solar cells [46]–[48].

Statistical Distribution of Shunt Conduction. Shunt current is of particular importance for module-level simulation because it is a key source of variability in individual cells. A recent study has shown that log-normal shunt distribution is universal in thin-film technologies (e.g., CIGS, CdTe, a-Si) [45]. Based on careful analysis of 34 commercial CIGS cells, we find that the log-normal distribution is justified here as well (Table A1 summarizes the key equations). The log-normal distribution exhibits a long "tail" in its probability density function. In a module, the highly shunted cells at the "tail" dissipate power generated by their neighboring good cells, degrading overall module efficiency [46].

The model in Fig. 1(c) is now fully specified, and the temperature/illumination dependencies are summarized in Table A1. We will now validate the model against experimental data.

### III. EXPERIMENTAL SETUP AND VALIDATION OF THE MODEL

The experiments were based on the standard high-efficiency (~18%) CIGS samples fabricated at the National Renewable Energy Laboratory (NREL). The CIGS absorbers were prepared by a co-evaporation method based on the three-stage process, followed by the deposition of CdS and ZnO layers and Ni/Al grid lines on top of the absorbers [49]. The detailed fabrication process and characterization results of similar devices from NREL have been described in [50], [51].

The model validation involves four steps:
1. *Measurement*: The I-V characteristics of the samples were measured as a function of temperature (282 K to 362 K) and illumination intensities (0 sun to 1 sun).

2. *Calibration*: We first fit the dark and light I-V characteristics under 1-sun illumination at a ***single*** temperature (294 K) to Eqs. (8) and (10) using a nonlinear least-squares fitting algorithm ("lsqcurvefit" function in Matlab® [52]). The typical numerical range for the initial guess of each parameter is specified in Table A2, and the specific fitting procedure is discussed in Table A3. The series and shunt resistance were extracted from the dark I-V curve using PVanalyzer [43].

3. *Prediction*: We extrapolated the parameters obtained in step 2 (from I-V at 294 K and 1-sun illumination) to all other temperatures and illumination intensities by applying (A.5) to (A.10) in Table A1.

4. *Validation*: The extrapolated parameters from step 3 were used as input to the model to predict the I-V curve as a function of temperature and illumination, and the outcome was compared to the measurements.

As shown in Fig. 2, the model predicts the salient features of the experimental I-V characteristics at various $T$ and $IL$ remarkably well. The validation suggests a noteworthy fact that *a single I-V measurement at room temperature and 1-sun intensity may be sufficient to predict cell response at arbitrary combinations of T and IL*, which is necessary to model cell-to-module gap, modules under partial shade, and lifetime energy output. Additionally, final fitting parameters in Eqs. (8) and (10) are physically relevant, and they provide insight

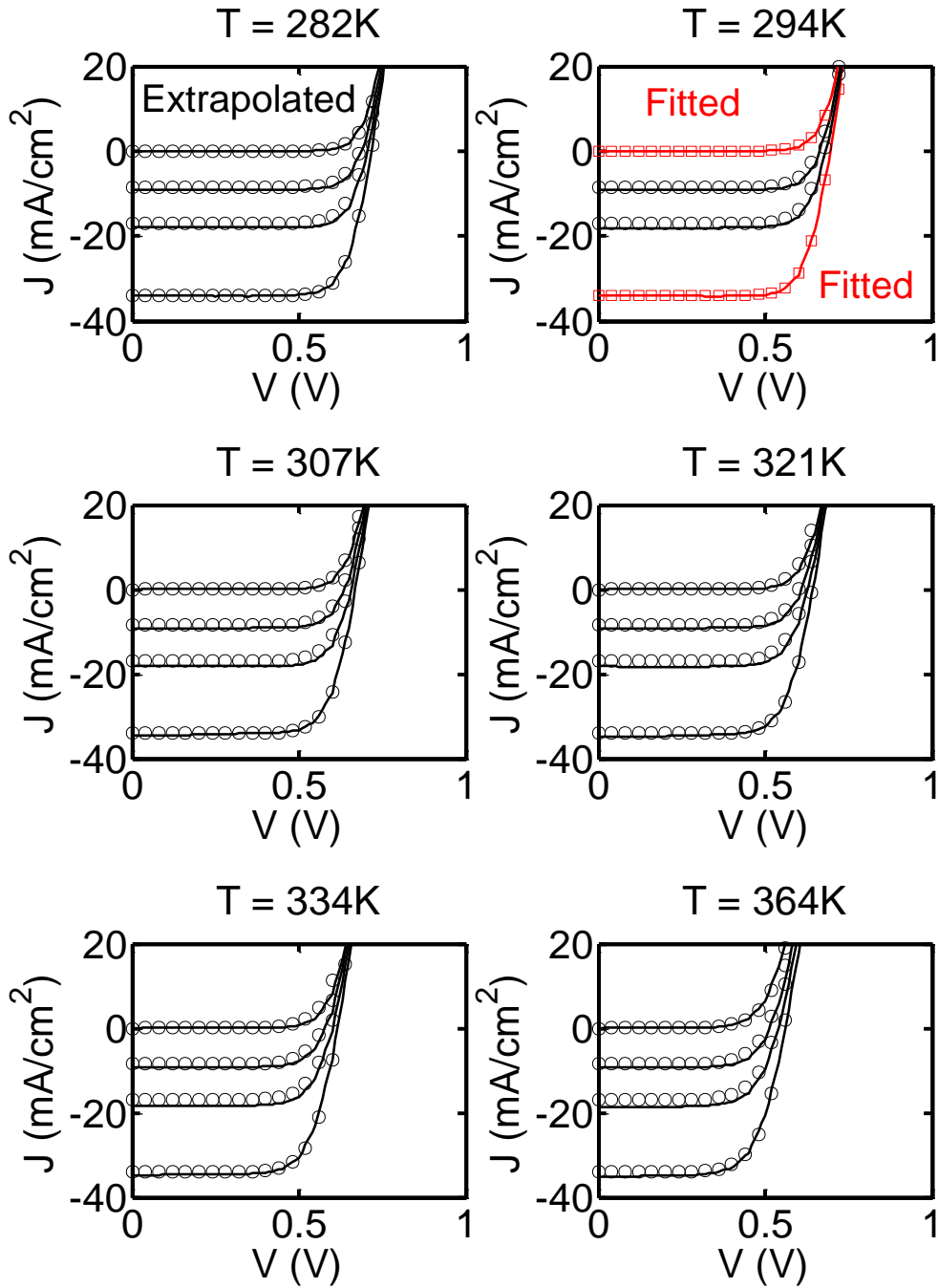

Fig. 2 The measured I-V (solid lines) vs. the analytical model (symbols) at different temperatures (282 K – 364 K) and illumination intensities (0 sun, 0.25 sun, 0.5 sun, and 1 sun). The square symbols denote the fitted data, whereas the circle symbols are all extrapolated results.

into the construction or pathology of the cells under measurement.

## IV. DISCUSSION AND CONCLUSIONS

<u>Advantages of a Physics-Based CIGS Model</u>. A physics-based model is powerful because it makes the following important contributions: *First*, the model can be used as a characterization technique. For example, among the parameters obtained by fitting the room-temperature data in Fig. 2, we found $V_{bi} \approx 0.8$ V (in good agreement with the C-V measurement). In additional, $E_G = 1.1$ eV, consistent with external quantum efficiency (EQE) measurement, and the heterojunction discontinuity for electrons $\Delta E_C = 0.1$ eV were used to extrapolate the I-V data at one sun and 298 K to arbitrary temperatures and illumination intensities. Hence, the model provides a means to calibrate physical parameters without performing C-V or EQE measurements. Indeed, parameters such as $\Delta E_C$ are critical, but are difficult to measure in any other way.

*Second*, the model allows us to predict efficiency at different temperatures and the temperature coefficient of the maximum output power without performing temperature-dependent light I-V measurement. Excellent agreement is obtained between two sets of experimental data and the analytical model; see Fig. 3 and Table 1. Note that, unlike [53], there are no empirical fitting parameters to account for the $T$ and $IL$ dependencies in this work and the model does not necessarily need to be calibrated against $T$-$IL$ data. Therefore, this model can be directly incorporated into a system-level simulation framework to physically predict the performance of CIGS-based solar modules under different ambient temperature and solar irradiance. Together with the model of module lifetime from [54], the simulation framework can also estimate the long-term energy gain of CIGS solar modules for various geographic locations and weather conditions, which provides useful projection and guidance for large-scale PV installations.

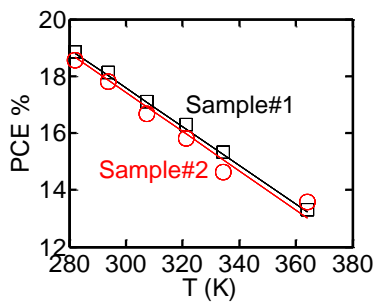

Fig. 3 Power conversion efficiency (PCE) of sample#1 (square) and sample#2 (circle) vs. the analytical model (solid and dashed lines for #1 and #2, respectively) as a function of temperature.

TABLE I. Measured and simulated temperature coefficients with uncertainties for $P_{max}$ of the two NREL samples.

| $P_{max}$ Temp. Coeff. (%K$^{-1}$) | Sample #1 | Sample #2 |
|---|---|---|
| Measurement | $-0.38 \pm 0.04$ | $-0.40 \pm 0.05$ |
| Analytical Model | $-0.38 \pm 0.04$ | $-0.39 \pm 0.07$ |

*Third* and finally, the model supports the development of an electro-thermal coupled simulation framework for solar modules. Equivalent circuits based on this model, which can accurately describe the temperature-dependent I-V characteristics at the cell level, can be integrated into a large-scale circuit network to self-consistently simulate the internal electrical and thermal distribution of a module. Such an application has been demonstrated to model shadow-induced self-heating in CIGS modules [6], and the simulation framework is also transferrable to interpret thermographic imaging for characterizing solar modules [55]. This paper focuses exclusively on the forward-bias response of CIGS cells; however, to properly simulate shadow-induced degradation at the module level, it is also necessary to analytically model the light-enhanced reverse-breakdown characteristics in CIGS [56]. We have also derived a semi-empirical model of the reverse-breakdown I-V, which can be found in [6].

<u>Model Limitations</u>. The model, nevertheless, has its limitations. For highly defective CIGS cells, several assumptions in the derivation, e.g., illumination-independent diode (injection) current, may not be valid [14] and the assumption of voltage-independent generation-induced bulk recombination current is not accurate for low-doped absorbers with $W(V) \approx t$. In addition, the bias- and light-induced metastability of defect response is not considered in the process of assessing the temperature and illumination dependencies [57], [58]. Therefore, the model must be used very carefully for very low-efficiency cells, because it is developed based on CIGS cells with moderately high efficiency.

To summarize, in this paper, we have presented a physics-based analytical model for p-n heterojunction solar cells, specifically, CIGS. The model has been validated against experimental data and has demonstrated the capability of describing I-V characteristics as a function of temperature and illumination with a few physical parameters. The log-normal distribution of the non-ohmic shunt leakage current is also included in the model to account for cell-level variability within solar modules. The main contributions of the model are as follows:

1. The model provides a simple recipe to characterize the physical parameters (e.g., $V_{bi}$) of CIGS solar cell only from I-V data and to estimate the temperature/illumination coefficients of efficiency without performing temperature-and illumination-dependent measurements.

2. The model can be integrated into an electro-thermal coupled module simulation framework to further investigate and reduce the cell-to-module efficiency gap [3], improve module reliability [6], as well as interpret thermal imaging measurements of solar modules [55].

3. At the system level, given the climatic and geographic information, the model can be used to accurately predict the long-term electricity yields for large-scale PV farms, which can provide guidance on solar installation for a given location.

ACKNOWLEDGMENT

This work was supported by the U.S. Department of Energy under Contract No. DE-AC36-08GO28308 with the National Renewable Energy Laboratory, the U.S. Department of Energy under DOE Cooperative Agreement no. DE-EE0004946 ("PVMI Bay Area PV Consortium"), the National Science Foundation through the NCN-NEEDS program, contract 1227020-EEC, and by the Semiconductor Research Corporation. The authors would like to thank Dr. Jian V. Li and Dr. Miguel A. Contreras for the measurement, Raghu V. Chavali and Dr. Mohammad R. Khan for helpful discussion, and Prof. Mark S. Lundstrom, Dr. Ingrid Repins, and Dr. Sarah.Kurtz for kind guidance.

## APPENDIX A (ANALYTICAL EQUATIONS AND FITTING/EXTRAPOLATION PROCEDURE)

First, we present the analytical descriptions of the model, the definition of each parameter as well as their illumination and temperature dependencies, and the statistical equation of the shunt distribution in Tables A1 and A2. The fitting and extrapolation procedures of measured I-V data are also specified in Table A3.

TABLE A1. The equations of the analytical model

| Analytical equations for I-V characteristics | |
|---|---|
| $J_{Photo} = J_{SC} \dfrac{1}{1 + \alpha_c e^{\frac{q\beta(V-V_{bi})}{kT}}}$ | (A.1) |
| $J_{Diode} = J_0(e^{\frac{qV}{NkT}} - 1)$ | (A.2) |
| $J_{Shunt} = G_{Shunt} \times V + I_{OShunt} \times V^\gamma$ | (A.3) |
| $J_{Light} = J_{Photo} + J_{Diode} + J_{Shunt}$ | (A.4) |
| Illumination and temperature dependencies of the parameters | |
| $J_{SC}$ | $J_{SC,IL} = IL \times J_{SC,one\ sun}$ (A.5) |
| $\alpha_c$ | $\alpha_{c,T} = \alpha_{c,300K} \times \exp\left(\dfrac{\Delta E_C}{k}\left(\dfrac{1}{300K} - \dfrac{1}{T}\right)\right)$ (A.6) |
| $\beta$ | $\beta_T = \beta_{300K}$ (A.7) |
| $V_{Bi}$ | $V_{Bi,T} = \dfrac{\Delta E_C}{q} + \dfrac{E_G}{q} + \dfrac{T}{300K}\left(V_{Bi,300K} - \dfrac{\Delta E_C}{q} - \dfrac{E_G}{q}\right) - \dfrac{kT}{q}\log\left(\left(\dfrac{T}{300K}\right)^3\right)$ (A.8) |
| $N$ | $N_T = 2/(1 + \dfrac{T}{T^*})$ (A.9) |
| $J_0$ | $J_{0,T} = J_{0,300\ K} \times \exp\left(\dfrac{E_G}{k}\left(\dfrac{1}{N_{300\ K}300\ K} - \dfrac{1}{N_T T}\right)\right)$ (A.10) |
| Statistical equation for log-normal shunt distribution | |
| $PDF(I_{OShunt}\|\mu,\sigma)$ | $\dfrac{1}{I_{OShunt}\sigma\sqrt{2\pi}} \times \exp\left(\dfrac{-(\log(I_{OShunt}) - \mu)^2}{2\sigma^2}\right)$ (A.11) |

\* $IL$ in (A.5) is the illumination intensity normalized to one sun, e.g., $IL = 0.2$ under 0.2-sun illumination.
\*\* In (A.11), $\mu$ and $\sigma$ are the $log$ mean and standard deviation, respectively. The actual mean and standard deviation of $G_{shunt}$ are $\exp(\mu + \dfrac{\sigma^2}{2})$ and $sqrt(\exp(2\mu + \sigma^2)/(\exp(\sigma^2) - 1))$, respectively.
\*\*\* Note that using different $E_G$ in (A.8) and (A.10) may provide a better benchmark against experimental data depending on samples.

TABLE A2. Definition and typical values—at room temperature (300 K) and one sun—for the parameters.

| Fitting parameters for dark I-V | | |
|---|---|---|
| $J_0$ | Diode saturation Current | ~ $10^{-6}$ mA/cm$^2$ |
| $N$ | Ideality factor | 1–2 |
| $G_{Shunt}$ | Prefactor of linear shunt current | 0–1 mS/cm$^2$ |
| $\gamma$ | Power index of log shunt current | 2–3 |
| $I_{OShunt}$ | Prefactor of nonlinear shunt current | 0–1 mS/(V$^{\gamma-1}$ cm$^2$) |
| Fitting parameters for $J_{Photo} = J_{Light} - J_{Diode}$ | | |
| $J_{sc}$ | Short-circuit current | 35 mA/cm$^2$ |
| $V_{Bi}$ | Total built-in voltage of the p-n junction | 0.6–0.9 V |
| $\alpha_c$ | Ratio between diffusion velocity and thermionic-emission velocity | 50–200 |
| Parameters used for extrapolation to arbitrary temperatures | | |
| $kT^*$ | Characteristic slope of the bulk defects | 60 ~ 200 meV |
| $E_G$ | Bandgap of CIGS | 1.0 ~ 1.4 eV |
| $\Delta E_C$ | Conduction band offset between CdS and CIGS layers | 0.1 eV |

TABLE A3. Fitting and extrapolation procedures of the model

| Fitting Procedure | |
|---|---|
| Step 1 ($J_{Shunt}$) | Under reverse biases, the dark current is dominated by the shunt current. Hence, by fitting the model (Eq. (A.3)) to the reverse-bias region (-1 V to 0 V) of dark I-V, one can extract the parameters for shunt current ($G_{Shunt}, \gamma$, and $I_{Oshunt}$). |
| Step 2 ($J_{Diode}$) | The next thing to do is to "clean" the dark IV. Due to the symmetric characteristics of shunt current ($J_{Shunt}(V) = J_{Shunt}(-V)$), one can separate the diode current and shunt current by subtracting $J_{Dark}(V)$ in the forward bias by its counterpart $J_{Dark}(-V)$ in the reverse bias (i.e., $J_{Diode}(V) = J_{Dark}(V) - J_{Dark}(-V)$). By applying Eq. (A.2) to the "cleaned" dark I-V, one can extract the parameters for diode current ($J_0$ and $N$). Steps 1–2 can be performed using the online tool PVanalyzer [43]. |
| Step 3 ($J_{Photo}$) | The photocurrent, $J_{Photo}$, is the difference between $J_{Light}$ and $J_{Dark}$ given that the diode current is illumination-independent. Therefore, one can extract $J_{Photo} = J_{Light} - J_{Dark}$ and fit it with (A.1) to obtain $J_{sc}, V_{Bi}$, and $\alpha_c$. |
| Extrapolation Procedure | |
| After extracting the parameters for a single temperature and illumination intensity, one can apply (A.5) to (A.11) to extrapolate I-V to arbitrary temperatures and illuminations. Note that those extrapolation parameters ($kT^*, E_G$ and $\Delta E_C$) may have a small variation between different cells. One can either calibrate these parameters if temperature-dependent I-V measurement is given, or use the typical values in Table A2 to predict the temperature coefficient of efficiency without performing the I-V measurement at different temperatures. | |

APPENDIX B (ASSUMPTION VERIFICATION)

Here, we validate the assumptions used to derive the analytical equations based on simulation and experiment data. The simulation was performed using the commercial device simulator, Sentaurus [22]. The discussion is divided into two parts: 1) illumination-independent diode current, and 2) voltage-independent generation-induced bulk recombination current.

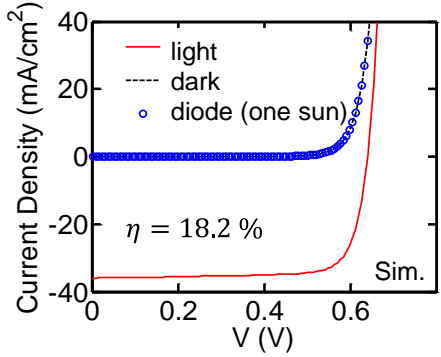

Fig. A1 The simulated I-V characteristics with 18.2% efficiency, showing that the frozen potential diode (injection) current is the same as dark current.

*A. Illumination-Independent Diode Current*

We simulate a CIGS solar cell with 18.2% efficiency; see Fig. A1. The device parameters are summarized in Appendix C. We take the approach discussed in [21] to separate the diode current and photocurrent under illumination, where the illuminated diode (injection) current was obtained by freezing the solution of the potential profile obtained under illumination and solving the frozen potential drift-diffusion equations after setting the photo-generation to zero. As shown in Fig. A1, $J_{Diode}(one\ sun) = J_{Dark}$ is found for $V < V_{OC}$. We have also experimentally proved that the diode current is illumination-independent by calculating the carrier-collection efficiencies ($\eta(V) = (J_{Light}(V) - J_{Dark}(V))/J_{SC}$) for different illumination intensities; see Fig. A2 (note that it is the same CIGS sample as shown in Fig. 2). The argument is that if $J_{Diode} = J_{Dark}$ under different sun intensities, the collection efficiency should not vary with illumination [17]. The results in Fig. A2 show that the collection efficiencies overlap each other for different illumination intensities (1 sun, 0.8 sun, 0.6 sun, 0.5 sun, 0.4 sun) at various temperatures except for $T = 364$ K, possibly due to high-level injection at high temperature. Therefore, based on the experimental analysis along with the simulation results, it is convincing that the diode current is illumination-independent for high-quality CIGS solar cells.

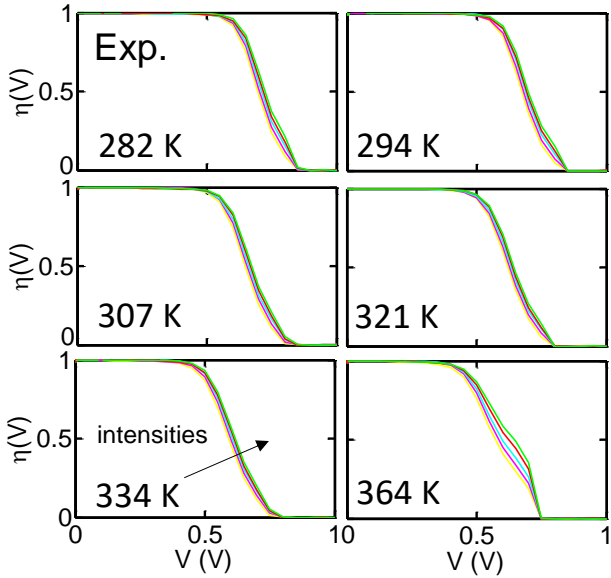

Fig. A2 The extracted collection efficiency, $\eta(V)$, under varying illumination intensity (1 sun, 0.8 sun, 0.6 sun, 0.5 sun, 0.4 sun) at different temperatures. The plots confirm the assumption of illumination-independent diode current except for 364 K.

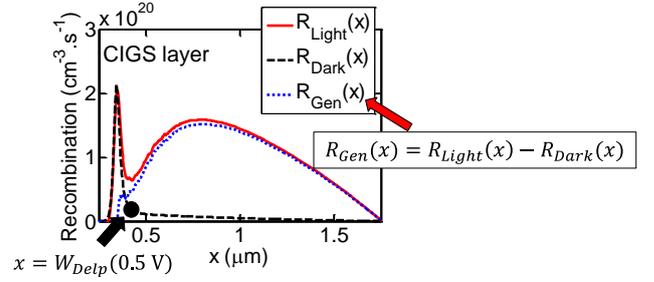

Fig. A3 The bulk recombination rate as a function of position inside the CIGS layer at 0.5 V. The generation-induced bulk recombination (dots) occurs mostly in the quasi-neutral region.

*B. Voltage-Independent Generation-Induced Bulk Recombination Current*

Here, we try to validate the assumption that the generation-induced bulk recombination current ($J_{Gen-Rec} = \int_o^t (R_{Light}(x) - R_{Dark}(x))dx = \int_o^t R_{Gen}(x)dx$) is voltage-independent up to $V_{OC}$ through simulation. The significance of the voltage-independent generation-induced bulk recombination current is that it allows us to simplify the term "$G(x) - R_{Gen}(x)$" in Eq. (2) to a normalize generation profile, "$G'(x)$", as discussed in Sec. II.

The bulk recombination rates under dark ($R_{Dark}(x)$) and light ($R_{Light}(x)$) and the generation-induced bulk recombination ($R_{Gen}(x)$) are plotted in Fig. A3. At $V = 0.5$ V, although there are considerable $R_{Dark}(x)$ and $R_{Light}(x)$ occurring in the depletion region ($x < 0.4$ μm), $R_{Gen}(x)$ occurs mostly in the quasi-neutral bulk region. Hence, $R_{Gen}(x)$ is voltage-independent. Consequently, as shown in Fig. A4(a), the generation-induced bulk recombination current, $J_{Gen-Rec}$, is a small portion of the total generation current and exhibits voltage-independent characteristics up to $V_{OC}$. Beyond $V_{OC}$, $J_{Gen-Rec}$ increases with voltage due to high-level injection. In addition, $J_{Gen-Rec}$ remains at 5% of the total generation current, $J_{Tot-Photo}$, for different illumination; see Fig. A4(b). So the short-circuit current (the maximum photocurrent), $J_{SC}$, can be written as $J_{Tot-Photo} - J_{Gen-Rec} = 0.95 \times J_{Tot-Photo}$ and is linear with illumination intensity.

So far, it has been confirmed that the diode current is voltage-independent, and the generation-induced bulk recombination current is voltage-independent under normal operating conditions. With these two assumptions, we can derive Eqs. (5)–(7) for photocurrent.

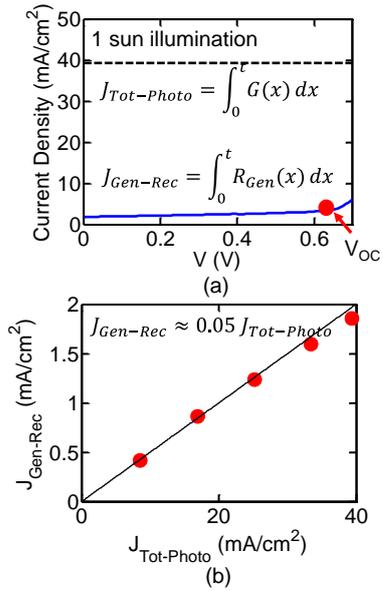

Fig. A4 (a) Total photocurrent ($J_{Tot-Photo}$) and generation-induced bulk recombination current ($J_{Gen-Rec}$) as a function of bias under one-sun conditions. $J_{Gen-Rec}$ is a small portion of $J_{Gen-Rec}$ and remains almost voltage-independent up to open-circuit voltage. (b) The linear relation between $J_{Gen-Rec}$ and $J_{Tot-Photo}$ for different illumination intensities.

APPENDIX C (DEVICE SIMULATION PARAMETERS)

TABLE A3. The simulation parameters used in Sentaurus numerical simulations [23]

| | Properties | n-layer | n-layer | p-layer |
|---|---|---|---|---|
| | Thickness | 200 nm | 50 nm | 1.5 µm |
| | Doping (cm$^{-3}$) | $N_D = 1 \times 10^{18}$ | $N_D = 1 \times 10^{17}$ | $N_A = 2 \times 10^{16}$ |
| | Hole mobility (cm$^2$/Vs) | 25 | 25 | 25 |
| | Electron mobility (cm$^2$/Vs) | 100 | 100 | 100 |
| | Bandgap (eV) | 3.3 | 2.4 | 1.15 |
| | Electron affinities (eV) | 4.4 | 4.2 | 4.5 |
| Defect | Type | Donor | Acceptor | Donor |
| | Defect level (eV) | Midgap | Midgap | Midgap |
| | Gaussian distribution width (eV) | 0.1 | 0.1 | 0.1 |
| | Defect density (cm$^{-3}$) | $1 \times 10^{17}$ | $1 \times 10^{17}$ | $2 \times 10^{14}$ |
| | Hole cross-section (cm$^2$) | $1 \times 10^{-12}$ | $1 \times 10^{-17}$ | $5 \times 10^{-13}$ |
| | Electron cross-section (cm$^2$) | $1 \times 10^{-15}$ | $1 \times 10^{-12}$ | $1 \times 10^{-15}$ |
| | Contact properties | Ohmic contacts, $s_f = 10^7 cm/s$ for both contacts | | |

REFERENCES

[1] "Solar Frontier Achieves World Record Thin-Film Solar Cell Efficiency: 22.3%," 2015. [Online]. Available: http://www.solar-frontier.com/eng/news/2015/C051171.html.

[2] M. A. Green, K. Emery, Y. Hishikawa, W. Warta, and E. D. Dunlop, "Solar cell efficiency tables (version 47)," *Prog. Photovoltaics Res. Appl.*, vol. 24, no. 1, pp. 3–11, Jan. 2016.

[3] S. Dongaonkar and M. A. Alam, "In-Line Post-Process Scribing for Reducing Cell to Module Efficiency Gap in Monolithic Thin-Film Photovoltaics," *IEEE J. Photovoltaics*, vol. 4, no. 1, pp. 324–332, Jan. 2014.

[4] E. S. Mungan, Y. Wang, S. Dongaonkar, D. R. Ely, R. E. García, and M. A. Alam, "From process to modules: End-to-end modeling of CSS-deposited CdTe solar cells," *IEEE J. Photovoltaics*, vol. 4, no. 3, pp. 954–961, 2014.

[5] M. A. Alam, B. Ray, M. R. Khan, and S. Dongaonkar, "The essence and efficiency limits of bulk-heterostructure organic solar cells: A polymer-to-panel perspective," *J. Mater. Res.*, vol. 28, no. 04, pp. 541–557, Feb. 2013.

[6] T. J. Silverman, M. G. Deceglie, X. Sun, R. L. Garris, M. A. Alam, C. Deline, and S. Kurtz, "Thermal and Electrical Effects of Partial Shade in Monolithic Thin-Film Photovoltaic Modules," *IEEE J. Photovoltaics*, vol. 5, no. 6, pp. 1742–1747, Nov. 2015.

[7] B. E. Pieters, "Spatial modeling of thin-film solar modules using the network simulation method and SPICE," *IEEE J. Photovoltaics*, vol. 1, no. 1, pp. 93–98, 2011.

[8] S. Dongaonkar, C. Deline, and M. A. Alam, "Performance and reliability implications of two-dimensional shading in monolithic thin-film photovoltaic modules," *IEEE J. Photovoltaics*, vol. 3, no. 4, pp. 1367–1375, 2013.

[9] J. Wong, "Griddler: Intelligent computer aided design of complex solar cell metallization patterns," in *2013 IEEE 39th Photovoltaic Specialists Conference (PVSC)*, 2013, pp. 0933–0938.

[10] M. Hejri, H. Mokhtari, M. R. Azizian, M. Ghandhari, and L. Soder, "On the Parameter Extraction of a Five-Parameter Double-Diode Model of Photovoltaic Cells and Modules," *IEEE J. Photovoltaics*, vol. 4, no. 3, pp. 915–923, May 2014.

[11] M. T. Boyd, S. a. Klein, D. T. Reindl, and B. P. Dougherty, "Evaluation and Validation of Equivalent Circuit Photovoltaic Solar Cell Performance Models," *J. Sol. Energy Eng.*, vol. 133, no. 2, p. 021005, 2011.

[12] X. Sun, R. Asadpour, W. Nie, A. D. Mohite, and M. A. Alam, "A Physics-Based Analytical Model for Perovskite Solar Cells," *IEEE J. Photovoltaics*, vol. 5, no. 5, pp. 1389–1394, Sep. 2015.

[13] S. Dongaonkar, X. Sun, M. Lundstrom, and M. A. Alam, "TAG Solar Cell Model (p-i-n thin film).," 2014. [Online]. Available: https://nanohub.org/publications/20.

[14] M. Gloeckler, C. R. Jenkins, and J. R. Sites, "Explanation of Light/Dark Superposition Failure in CIGS Solar Cells," *MRS Proc.*, vol. 763, p. B5.20, Jan. 2003.

[15] M. Eron and A. Rothwarf, "Effects of a voltage-dependent light-generated current on solar cell measurements: CuInSe2/Cd(Zn)S," *Appl. Phys. Lett.*, vol. 44, no. 1, p. 131, 1984.

[16] J. E. Moore, S. Dongaonkar, R. V. K. Chavali, M. A. Alam, and M. S. Lundstrom, "Correlation of built-in potential and I-V crossover in thin-film solar cells," *IEEE J. Photovoltaics*, vol. 4, no. 4, pp. 1138–1148, 2014.

[17] S. Hegedus, D. Desai, and C. Thompson, "Voltage dependent photocurrent collection in CdTe/CdS solar cells," *Prog. Photovoltaics Res. Appl.*, vol. 15, no. 7, pp. 587–602, Nov. 2007.

[18] X. X. Liu and J. R. Sites, "Solar-cell collection efficiency and its variation with voltage," *J. Appl. Phys.*, vol. 75, no. 1, pp. 577–581, Jan. 1994.

[19] K. W. Mitchell, A. L. Fahrenbruch, and R. H. Bube, "Evaluation of the CdS/CdTe heterojunction solar cell," *J. Appl. Phys.*, vol. 48, no. 10, pp. 4365–4371, 1977.

[20] W. W. Gärtner, "Depletion-layer photoeffects in semiconductors," *Phys. Rev.*, vol. 116, pp. 84–87, 1959.

[21] J. L. G. R. V. K. Chavali, J. E. Moore, X. Wang, M. A. Alam, M. S. Lundstrom, R. V. K. Chavali, J. E. Moore, X. Wang, M. A. Alam, M. S. Lundstrom, and J. L. Gray, "The Frozen Potential Approach to Separate the Photocurrent and Diode Injection Current in Solar Cells," *IEEE J. Photovoltaics*, vol. 5, no. 3, pp. 865–873, May 2015.

[22] "Synopsys Sentaurus Semiconductor TCAD Software." East Middlefield Road, Mountain View, CA 94043 USA.

[23] M. Gloeckler, a. L. Fahrenbruch, and J. R. Sites, "Numerical modeling of CIGS and CdTe solar cells: setting the baseline," *3rd World Conf. onPhotovoltaic Energy*

[24] R. F. Pierret, *Semiconductor Device Fundamentals*. Prentice Hall, 1996.
[25] S. M. Sze and K. K. Ng, *Physics of Semiconductor Devices*, 3 Edition. Wiley-Interscience, 2006.
[26] J. Gray, X. Wang, R. Chavali, X. Sun, K. Abhirit, and J. Wilcox, "ADEPT 2.1," 2015. [Online]. Available: https://nanohub.org/resources/adeptnpt.
[27] Y. Taur and T. H. Ning, *Fundamentals of Modern VLSI Devices*, 2nd Ed. Cambridge University Press, 2009.
[28] T. Song, J. Tyler McGoffin, and J. R. Sites, "Interface-Barrier-Induced J–V Distortion of CIGS Cells With Sputtered-Deposited Zn(S,O) Window Layers," *IEEE J. Photovoltaics*, vol. 4, no. 3, pp. 942–947, May 2014.
[29] P. E. Russell, O. Jamjoum, R. K. Ahrenkiel, L. L. Kazmerski, R. a. Mickelsen, and W. S. Chen, "Properties of the Mo-CuInSe2 interface," *Appl. Phys. Lett.*, vol. 40, no. 11, pp. 995–997, 1982.
[30] X. Sun, C. J. Hages, N. J. Carter, J. E. Moore, R. Agrawal, and M. Lundstrom, "Characterization of nanocrystal-ink based CZTSSe and CIGSSe solar cells using voltage-dependent admittance spectroscopy," in *2014 IEEE 40th Photovoltaic Specialist Conference (PVSC)*, 2014, vol. 2, pp. 2416–2418.
[31] T. Eisenbarth, T. Unold, R. Caballero, C. a. Kaufmann, and H.-W. Schock, "Interpretation of admittance, capacitance-voltage, and current-voltage signatures in Cu(In,Ga)Se[sub 2] thin film solar cells," *J. Appl. Phys.*, vol. 107, no. 3, p. 034509, 2010.
[32] S. H. Demtsu and J. R. Sites, "Effect of back-contact barrier on thin-film CdTe solar cells," *Thin Solid Films*, vol. 510, no. 1–2, pp. 320–324, Jul. 2006.
[33] R. V. K. Chavali, J. R. Wilcox, B. Ray, J. L. Gray, and M. A. Alam, "Correlated nonideal effects of dark and light I–V characteristics in a-Si/c-Si heterojunction solar cells," *IEEE J. Photovoltaics*, vol. 4, no. 3, pp. 763–771, May 2014.
[34] S. S. Hegedus and W. N. Shafarman, "Thin-film solar cells: device measurements and analysis," *Prog. Photovoltaics Res. Appl.*, vol. 12, no. 23, pp. 155–176, Mar. 2004.
[35] C. Rincón, S. M. Wasim, G. Marín, and I. Molina, "Temperature dependence of the optical energy band gap in CuIn3Se5 and CuGa3Se5," *J. Appl. Phys.*, vol. 93, no. 1, p. 780, 2003.
[36] "Miasole: Datasheet FLEX-02N." [Online]. Available: http://miasole.com/wp-content/uploads/2015/08/FLEX-02N_Datasheet_1.pdf.
[37] M. A. Contreras, J. Tuttle, A. Gabor, A. Tennant, K. Ramanathan, S. Asher, A. Franz, J. Keane, L. Wang, J. Scofield, and R. Noufi, "High efficiency Cu(In,Ga)Se2-based solar cells: processing of novel absorber structures," in *Proceedings of 1994 IEEE 1st World Conference on Photovoltaic Energy Conversion - WCPEC (A Joint Conference of PVSC, PVSEC and PSEC)*, vol. 1, pp. 68–75.
[38] M. S. Lundstrom, "ECE 612 Lecture 31: Heterostructure Fundamentals," 2006. [Online]. Available: https://nanohub.org/resources/2082.
[39] U. Rau, a. Jasenek, H. W. Schock, F. Engelhardt, and T. Meyer, "Electronic loss mechanisms in chalcopyrite based heterojunction solar cells," *Thin Solid Films*, vol. 361, pp. 298–302, 2000.
[40] T. Walter, R. Herberholz, and H.-W. Schock, "Distribution of Defects in Polycrytalline Chalcopyrite Thin Films," *Solid State Phenom.*, vol. 51–52, pp. 309–316, 1996.
[41] S. Dongaonkar and M. A. Alam, "Geometrical design of thin film photovoltaic modules for improved shade tolerance and performance," *Progress in Photovoltaics: Research and Applications*, vol. 20, no. 1, pp. 6–11, 2013.
[42] G. T. Koishiyev and J. R. Sites, "Impact of sheet resistance on 2-D modeling of thin-film solar cells," *Sol. Energy Mater. Sol. Cells*, vol. 93, no. 3, pp. 350–354, Mar. 2009.
[43] S. Dongaonkar and M. A. Alam, "PV Analyzer," 2014. .
[44] S. Dongaonkar, Y. Karthik, S. Mahapatra, and M. A. Alam, "Physics and statistics of non-ohmic shunt conduction and metastability in amorphous silicon p-i-n solar cells," *IEEE J. Photovoltaics*, vol. 1, no. 2, pp. 111–117, 2011.
[45] S. Dongaonkar, J. D. Servaites, G. M. Ford, S. Loser, J. Moore, R. M. Gelfand, H. Mohseni, H. W. Hillhouse, R. Agrawal, M. A. Ratner, T. J. Marks, M. S. Lundstrom, and M. A. Alam, "Universality of non-Ohmic shunt leakage in thin-film solar cells," *J. Appl. Phys.*, vol. 108, no. 12, p. 124509, 2010.
[46] S. Dongaonkar, S. Loser, E. J. Sheets, K. Zaunbrecher, R. Agrawal, T. J. Marks, and M. a. Alam, "Universal statistics of parasitic shunt formation in solar cells, and its implications for cell to module efficiency gap," *Energy Environ. Sci.*, vol. 6, no. 3, pp. 782–787, 2013.
[47] B. L. Williams, S. Smit, B. J. Kniknie, K. J. Bakker, W. Keuning, W. M. M. Kessels, R. E. I. Schropp, and M. Creatore, "Identifying parasitic current pathways in CIGS solar cells by modelling dark J-V response," *Prog. Photovoltaics Res. Appl.*, vol. 23, no. 11, pp. 1516–1525, Nov. 2015.
[48] B. J. Mueller, C. Zimmermann, V. Haug, F. Hergert, T. Koehler, S. Zweigart, and U. Herr, "Influence of different sulfur to selenium ratios on the structural and electronic properties of Cu(In,Ga)(S,Se)2 thin films and solar cells formed by the stacked elemental layer process," *J. Appl. Phys.*, vol. 116, no. 17, 2014.
[49] P. M. P. Salomé, V. Fjällström, P. Szaniawski, J. P. Leitão, A. Hultqvist, P. A. Fernandes, J. P. Teixeira, B. P. Falcão, U. Zimmermann, A. F. da Cunha, and M. Edoff, "A comparison between thin film solar cells made from co-evaporated CuIn 1-x Ga x Se 2 using a one-stage process versus a three-stage process," *Prog. Photovoltaics Res. Appl.*, vol. 23, no. 4, pp. 470–478, Apr. 2015.
[50] K. Ramanathan, M. A. Contreras, C. L. Perkins, S. Asher, F. S. Hasoon, J. Keane, D. Young, M. Romero, W. Metzger, R. Noufi, J. Ward, and A. Duda, "Properties of 19.2% efficiency ZnO/CdS/CuInGaSe2 thin-film solar cells," *Prog. Photovoltaics Res. Appl.*, vol. 11, no. 4, pp. 225–230, 2003.
[51] M. A. Contreras, M. J. Romero, B. To, F. Hasoon, R. Noufi, S. Ward, and K. Ramanathan, "Optimization of CBD CdS process in high-efficiency Cu(In,Ga)Se2-based solar cells," *Thin Solid Films*, vol. 403–404, pp. 204–211, Feb. 2002.
[52] "Matlab2014a." The MathWorks Inc, Natick, MA, 2014.
[53] W. De Soto, S. A. Klein, and W. A. Beckman, "Improvement and validation of a model for photovoltaic array performance," *Sol. Energy*, vol. 80, no. 1, pp. 78–88, Jan. 2006.
[54] P. Hacke, S. Spataru, K. Terwilliger, G. Perrin, S. Glick, S. Kurtz, and J. Wohlgemuth, "Accelerated Testing and Modeling of Potential-Induced Degradation as a Function of Temperature and Relative Humidity," *IEEE J. Photovoltaics*, vol. 5, no. 6, pp. 1549–1553, Nov. 2015.
[55] A. Gerber, V. Huhn, T. M. H. Tran, M. Siegloch, Y. Augarten, B. E. Pieters, and U. Rau, "Advanced large area characterization of thin-film solar modules by electroluminescence and thermography imaging techniques," *Sol. Energy Mater. Sol. Cells*, vol. 135, pp. 35–42, Apr. 2015.
[56] P. Szaniawski, J. Lindahl, T. Törndahl, U. Zimmermann, and M. Edoff, "Light-enhanced reverse breakdown in Cu(In,Ga)Se2 solar cells," *Thin Solid Films*, vol. 535, pp. 326–330, May 2013.

[Conversion, 2003. Proc.*, vol. 1, pp. 491–494, 2003.]

[57] S. Lany and A. Zunger, "Light- and bias-induced metastabilities in Cu(In,Ga)Se2 based solar cells caused by the (VSe-VCu) vacancy complex," *J. Appl. Phys.*, vol. 100, pp. 1–15, 2006.

[58] J. T. Heath, J. D. Cohen, and W. N. Shafarman, "Distinguishing metastable changes in bulk CIGS defect densities from interface effects," *Thin Solid Films*, vol. 431–432, no. 03, pp. 426–430, May 2003.

**Xingshu Sun** (S'13) received the B.S. degree from Purdue University, West Lafayette, IN, in 2012, where he is currently working toward the Ph.D. degree in electrical and computer engineering. His current research interests include device simulation and compact modeling for solar cells and nanoscale transistors and system-level simulation for photovoltaics.

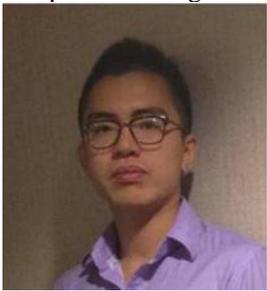

**Timothy Silverman** has a doctorate in mechanical engineering from the University of Texas at Austin (2010). He is an expert in heat transfer, solid mechanics, computational modeling, PV module reliability, and outdoor PV performance. At NREL, he performs research in the performance and reliability of silicon, polycrystalline thin film, and advanced concept PV modules.

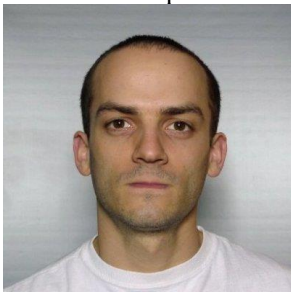

**Rebekah Garris** got her PhD in physics from the University of California at Santa Cruz. Her passion for using physics to help people flows through her work as a scientist at NREL. In addition, Rebekah enjoys giving back to the community as a physics and math teaching faculty member at several Denver-area colleges.

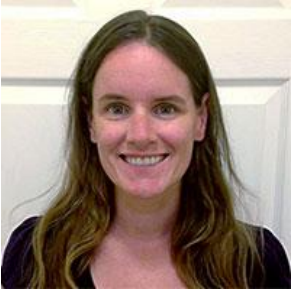

**Chris Deline** received the B.S., M.S., and Ph.D. degrees from the University of Michigan, Ann Arbor, in 2003, 2005, and 2008, respectively, all in electrical engineering. His Ph.D. work involved experimental investigation of the magnetic nozzle region of electric propulsion thrusters at NASA MSFC and JSC. Since 2008 he has been a research engineer at the National Renewable Energy Laboratory in Golden, CO in the photovoltaic performance and reliability group. He has co-authored several reports on the performance impact of microinverters and DC-DC converters in PV systems under mismatched and shaded conditions. His research includes PV field performance including thin-film PV performance and stabilization, mismatch and partial shading in PV systems and distributed power electronics for PV. Chris manages the DOE Regional Test Center program at NREL for field assessment of novel PV technologies, and is principal investigator for multiple projects on PV field performance including degradation rate assessment and bifacial module power rating and energy production modeling.

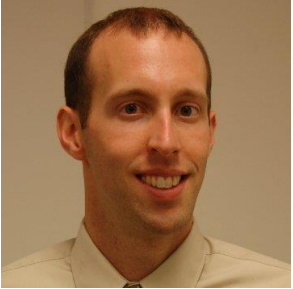

**Muhammad Ashraful Alam** (M'96–SM'01–F'06) is the Jai N. Gupta Professor of Electrical and Computer Engineering where his research and teaching focus on physics, simulation, characterization and technology of classical and emerging electronic devices. From 1995 to 2003, he was with Bell Laboratories, Murray Hill, NJ, where he made important contributions to reliability physics of electronic devices, MOCVD crystal growth, and performance limits of semiconductor lasers. At Purdue, Alam's research has broadened to include flexible electronics, solar cells, and nanobiosensors. He is a fellow of the AAAS, IEEE, and APS and recipient of the 2006 IEEE Kiyo Tomiyasu Award for contributions to device technology.

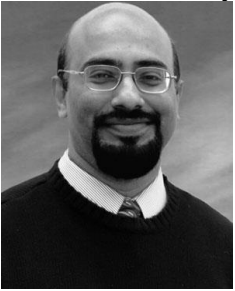